\documentclass[aps,prb,showpacs,twocolumn,showpacs]{revtex4}

\usepackage{graphicx}
\usepackage{amsmath}
\usepackage{amssymb}
\usepackage{color}
\usepackage[extra]{tipa}

\definecolor{lila}{rgb}{0.5,0,1}
\newcommand{\bnen}{\begin{equation}}
\newcommand{\eden}{\end{equation}}
\newcommand{\bean}{\begin{eqnarray}}
\newcommand{\eean}{\end{eqnarray}}

\newcommand{\bna}{\begin{array}}
\newcommand{\eda}{\end{array}}

\begin{document}

\title{Quantum criticality and first-order transitions in the extended periodic Anderson model}

\author{I. Hagym\'asi$^{1,2}$}
\author{K. Itai$^1$}
\author{J. S\'olyom$^{1}$}

\affiliation{$^1$Strongly Correlated Systems "Lend\"ulet" Research Group, Institute for Solid State Physics and Optics, MTA Wigner Research Centre for Physics, Budapest H-1525 P.O. Box 49, Hungary\\
$^2$Institute of Physics, E\"otv\"os University, Budapest, P\'azm\'any P\'eter s\'et\'any 1/A, 
H-1117, Hungary}

\date{\today}

\begin{abstract}
We investigate the behavior of the periodic Anderson model in the 
presence of $d$-$f$ Coulomb interaction ($U_{df}$) using mean-field theory, variational calculation, and exact diagonalization of finite chains. The variational approach based on the Gutzwiller trial wave function gives a critical value of $U_{df}$ and two quantum critical points (QCPs), where the valence susceptibility diverges. We derive the critical exponent for the valence susceptibility and investigate how the position of the QCP depends on the other parameters of the Hamiltonian. For larger values of $U_{df}$, the Kondo regime is bounded by two first-order transitions. These first-order transitions merge into a triple point at a certain value of  $U_{df}$. For even larger  $U_{df}$ valence skipping occurs. Although the other methods do not give a critical point, they support this scenario.

\end{abstract}

\pacs{71.10.Fd, 71.27.+a, 75.30.Mb}

\maketitle

\section{Introduction} 
Heavy-fermion compounds often show remarkable phenomena like unconventional superconductivity or unusual Fermi-liquid state. 
It turned out that in those compounds, for instance CeIn$_3$,\cite{Mathur:meres} whose superconducting state is unconventional, the pairing between electrons is mediated by antiferromagnetic spin fluctuations. The superconducting phase is formed near the antiferromagnetic quantum critical point (QCP)\cite{AF_SC:theory} as the pressure or in other cases the concentration of a component is varied. However, this theory seems to be insufficient to explain the temperature-pressure phase diagram of CeCu$_2$Ge$_2$ or CeCu$_2$Si$_2$, where a superconducting dome with enhanced transition temperature is located far away from the antiferromagnetic critical point. Since this discovery, these compounds have drawn much attention both experimentally and theoretically. It has been argued that this phenomenon is related to the critical valence fluctuations of Ce ions,\cite{CeCu2Ge2,Miyake:VMC,Miyake:CVF_1,Yuan:science,CeCu2SiGe:meres,DMRG:cikkek,Miyake:SC,CeCu2Si2:meres_2,Miyake:review,CeCu2Si2:meres_1,CeCu2Si2:meres_SC,RueffPRL_meres:cikk} that is, to the existence of a second QCP.
\par The simplest model of heavy-fermion compounds is the periodic Anderson model (PAM).\cite{Review} It is known, however, that 
the mixed-valence regime appears always in this model as a smooth crossover, and valence fluctuations do not become
critical for any choice of the parameters. A local Coulomb interaction between the conduction and localized electrons is
needed for the appearance of a sharp transition and critical valence fluctuations.\cite{Miyake:VMC,Miyake:review} The
Hamiltonian of this extended periodic Anderson model can be written using standard notations in the following form:
\begin{equation}   \begin{split}
     \mathcal{H} = & \sum_{\boldsymbol{k},\sigma}\varepsilon_d(\boldsymbol{k})
       \hat{d}_{\boldsymbol{k}\sigma}^{\dagger}
	   \hat{d}^{\phantom \dagger}_{\boldsymbol{k}\sigma}- V\sum_{\boldsymbol{j},\sigma}\left(\hat{f}_{\boldsymbol{j}\sigma}^{\dagger}
	  \hat{d}^{\phantom \dagger}_{\boldsymbol{j}\sigma}
    +\hat{d}_{\boldsymbol{j}\sigma}^{\dagger} \hat{f}^{\phantom \dagger}_{\boldsymbol{j}\sigma}\right)   \\\label{PAM:Hamiltonian}
	   &  +\varepsilon_f\sum_{\boldsymbol{j},\sigma}\hat{n}^f_{\boldsymbol{j}\sigma}
+U_f\sum_{\boldsymbol{j}}\hat{n}^f_{\boldsymbol{j}\uparrow}
\hat{n}^f_{\boldsymbol{j}\downarrow}+U_{df}\sum_{\boldsymbol{j},\sigma,\sigma'}
\hat{n}^f_{\boldsymbol{j}\sigma}\hat{n}^d_{\boldsymbol{j}\sigma'} . 
\end{split}
\end{equation}
After Onishi and Miyake's pioneering work,\cite{Miyake:VMC} recently, this model has been investigated by several modern techniques, including density matrix renormalization group,\cite{DMRG:cikkek} dynamical mean field,\cite{Kawakami:DMFT_1,Kawakami:DMFT_2,Hirashima:cikk} variational calculations,\cite{Miyake:VMC,Kubo:GW} projector based renormalization approach\cite{PBR:cikk}, and fluctuation exchange approximation.\cite{Hirashima:FLEX} It has been found that a first-order valence transition and a QCP may appear due to $U_{df}$.

\par Previous calculations\cite{Miyake:VMC,Hirashima:cikk,Kubo:GW} focused on the properties at infinite or large $U_f$.
Mainly the existence of a QCP and the possibility of first-order transition was addressed. 
Our main goal here is to study the critical behavior for arbitrary values of $U_f$. 
We investigate how the QCP and the $\varepsilon_f-U_{df}$ phase diagram depend on the parameters of the model in the
half-filled case. In our previous paper\cite{Hagymasi:long} we have shown that the Gutzwiller's variational method gives reliable
results concerning the valence, therefore it is worth studying the valence transition by this method. 
\par It is worth noting that a first-order transition from Mott insulator to Kondo insulator has been found\cite{Koga:cikk} in a model
with a more general Hamiltonian, too, including Hund's coupling and interaction between $d$-electrons. 
We do not consider the Hund's coupling here since the appearance of critical valence fluctuations was attributed to the direct Coulomb interaction between $d$- and
$f$-electrons.\cite{Miyake:VMC,Miyake:review} The exchange coupling between them probably plays a minor role in this respect.
\par Note that in a previous paper of ours\cite{Hagymasi:Acta_cikk} a variational approach was formulated for another kind of
extended periodic Anderson model, where a different form was chosen for the $d$-$f$ interaction, a spin-dependent four-body term. Neither a QCP, nor a first-order valence transition has been found in that model.
\par The setup of the paper is as follows. In Sec. II, we perform a mean-field
calculation to demonstrate in the simplest way how $U_{df}$ affects the
intermediate valence regime. In Sec. III, the variational approach is
introduced, which is based on the Gutzwiller wave function. We analyze the
quantum critical behavior and the disappearance of the Kondo regime at a triple
point.  Moreover, we construct the
$\varepsilon_f-U_{df}$ phase diagram. In Sec. IV, we
carry out exact diagonalization to investigate the model in one dimension and
compare the results with that of mean-field theory and the Gutzwiller approach. Finally, in Sec. V, our conclusions are presented.

\section{Mean-field calculation}
First of all we will study the problem by mean-field methods. In this approach some kind of order has to be assumed. We performed calculations by assuming two possibilities: a) the system is paramagnetic; b) it possesses a spatially oscillating magnetic order with a total magnetization zero. The state with lowest energy is accepted as the ground state. According to our calculations, the mean-field equations always have a paramagnetic solution, but in the Kondo regime, where localized moments are present, ordering of the moments leads to the lowering of the energy.
We assume a simple cubic lattice in the following calculations, which can be partitioned into two sublattices ($A$ and $B$). We expect that in the broken symmetry phase the electrons are ordered on the two sublattices in an alternating fashion, that is:
\begin{gather}
 \left\langle \hat{n}^f_{\boldsymbol{j}\sigma} \right\rangle=\frac{1}{2}\bigg[n_f+\sigma m_f e^{i\boldsymbol{q}_0\cdot\boldsymbol{R}_{j}}\bigg],
\end{gather}
where $m_f$ is the magnetization of the sublattice and $\boldsymbol{q}_0=\pi/a(1,1,1)$, so that $\boldsymbol{q}_0\boldsymbol{R}_j=2\pi n$ on sublattice $A$ and $\boldsymbol{q}_0\boldsymbol{R}_j=(2n+1)\pi$ on sublattice $B$ ($n$ is an integer), and $n_f$ is the average number of $f$-electrons per site. The values of $n_f$ and $m_f$ need to be determined self-consistently. Similar oscillation can be assumed for the $d$-electrons,
\begin{gather}
 \Big\langle \hat{n}^d_{\boldsymbol{j}\sigma} \Big\rangle=\frac{1}{2}\bigg[n_d+\sigma m_d e^{i\boldsymbol{q}_0\cdot\boldsymbol{R}_{j}}\bigg],
\end{gather}
although $m_d$ will not appear explicitly in the calculations.
The mean-field Hamiltonian is 
\begin{gather}
\mathcal{H}^{\rm m}_{\rm AF} =  \sum_{\boldsymbol{k},\sigma}\big[\varepsilon_d(\boldsymbol{k})+ U_{df}n_f\big] 
\hat{d}_{\boldsymbol{k}\sigma}^{\dagger}\hat{d}^{\phantom \dagger}_{\boldsymbol{k}\sigma} \nonumber\\
+\sum_{\boldsymbol{j},\sigma}\left[\varepsilon_f + \frac{U_f}{2}\left(n_f 
-\sigma e^{i\boldsymbol{q}_0\cdot\boldsymbol{R}_{j}}m_f\right) + U_{df}n_d\right] \hat{n}^f_{\boldsymbol{j}\sigma}\nonumber\\
 + V\sum_{\boldsymbol{j},\sigma}\left(\hat{f}_{\boldsymbol{j}\sigma}^{\dagger} \hat{d}^{\phantom \dagger}_{\boldsymbol{j}\sigma}+\hat{d}_{\boldsymbol{j}\sigma}^{\dagger} \hat{f}^{\phantom \dagger}_{\boldsymbol{j}\sigma}\right)\nonumber\\
- \frac{NU_f}{4} \left(n_f^2-m_f^2\right) -NU_{df}n_fn_d,\label{eq:AF}
\end{gather}
where the $\boldsymbol{k}$ sum extends over the whole Brillouin zone of the simple cubic lattice, and $N$ is the number of sites. Due to the assumed magnetic ordering, the size of the Brillouin zone is reduced to half of its original size. In order to restrict the $\boldsymbol{k}$ sum to the magnetic Brillouin zone, we split the original sum into two parts by introducing  the operators $\hat{d}_{\boldsymbol{k}+\boldsymbol{q_0}\sigma}^{\dagger}$ and $\hat{f}_{\boldsymbol{k}+\boldsymbol{q_0}\sigma}^{\dagger}$. We suppose that the dispersion relation possesses the nesting property:
\begin{gather} \label{eq:nesting}
 \varepsilon_d(\boldsymbol{k}+\boldsymbol{q}_0)=-\varepsilon_d(\boldsymbol{k}),
\end{gather}
which is valid in a tight-binding model with nearest neighbor hopping. This fixes the zero of the energy scale. Then the mean-field Hamiltonian can be rewritten in Bloch representation:
\begin{gather}
\mathcal{H}^{\rm m}_{\rm AF}=
\sum_{\boldsymbol{k},\sigma}{}^{'}
\left(\begin{array}{c}
\hat{d}_{\boldsymbol{k}\sigma}\\
\hat{d}_{\boldsymbol{k}+\boldsymbol{q_0}\sigma}\\
\hat{f}_{\boldsymbol{k}\sigma}\\
\hat{f}_{\boldsymbol{k}+\boldsymbol{q_0}\sigma}
\end{array}\right)^{\dagger}
\mathcal{H}(\boldsymbol{k},\sigma)
\left(\begin{array}{c}
\hat{d}_{\boldsymbol{k}\sigma}\\
\hat{d}_{\boldsymbol{k}+\boldsymbol{q_0}\sigma}\\
\hat{f}_{\boldsymbol{k}\sigma}\\
\hat{f}_{\boldsymbol{k}+\boldsymbol{q_0}\sigma}
\end{array}\right)\nonumber\\
- \frac{NU_f}{4} \left(n_f^2-m_f^2\right) -NU_{df}n_fn_d,
\end{gather}
where the prime denotes that the summation is carried out over the magnetic Brillouin zone, and
\begin{gather}
 \mathcal{H}(\boldsymbol{k},\sigma)=\left(\begin{array}{cccc}
\xi_d(\boldsymbol{k})&0&V&0\\
0&\tilde{\xi}_d(\boldsymbol{k})&0&V\\
V&0&\xi_f&-U_f\sigma m_f/2\\
0&V&-U_f\sigma m_f/2&\xi_f
\end{array}\right),
\end{gather}
where $\xi_f=\varepsilon_f + U_fn_f/2 + U_{df}n_d$,
$\xi_d(\boldsymbol{k})=\varepsilon_d(\boldsymbol{k})+U_{df}n_f$
and
$\tilde{\xi}_d(\boldsymbol{k})=-\varepsilon_d(\boldsymbol{k})+U_
{df}n_f$.
 This Hamiltonian can be diagonalized by the unitary transformation $\mathcal{T}(\boldsymbol{k}\sigma)$, which is a real matrix in our case, leading to
\begin{gather}
\mathcal{H}^{\rm m}_{\rm AF}  = \sum_{\boldsymbol{k},\sigma}{}^{'}\left[
E_a(\boldsymbol{k})\hat A_{\boldsymbol{k}\sigma}^{\dagger}\hat{A}_{\boldsymbol{k}\sigma}
+E_b(\boldsymbol{k})\hat B_{\boldsymbol{k}\sigma}^{\dagger}\hat{B}_{\boldsymbol{k}\sigma}\right.\nonumber\\
\left.+E_c(\boldsymbol{k})\hat C_{\boldsymbol{k}\sigma}^{\dagger}\hat{C}_{\boldsymbol{k}\sigma}
+E_d(\boldsymbol{k})\hat D_{\boldsymbol{k}\sigma}^{\dagger}\hat{D}_{\boldsymbol{k}\sigma}\right]\nonumber\\
- \frac{NU_f}{4} \left(n_f^2-m_f^2\right) -NU_{df}n_fn_d,
\end{gather}
where
\begin{gather}
\Big(\begin{array}{cccc}
\hat{A}_{\boldsymbol{k}\sigma}^{\dagger}&\hat{B}_{\boldsymbol{k}\sigma}^{\dagger}&\hat{C}_{\boldsymbol{k}\sigma}^{\dagger}&\hat{D}_{\boldsymbol{k}\sigma}^{\dagger}
\end{array}\Big)\nonumber\\
=
\left(\begin{array}{cccc}
\hat{d}_{\boldsymbol{k}\sigma}^{\dagger}&\hat{d}_{\boldsymbol{k}+\boldsymbol{q}_0\sigma}^{\dagger}&\hat{f}_{\boldsymbol{k}\sigma}^{\dagger}&\hat{f}_{\boldsymbol{k}+\boldsymbol{q}_0\sigma}^{\dagger}
\end{array} \right)\mathcal{T}^{\dagger}(\boldsymbol{k}\sigma).
\end{gather}
The diagonalization is done numerically for each $\boldsymbol{k}$ value and we sort the eigenvalues in increasing order $\left[E_a(\boldsymbol{k})\leq E_b(\boldsymbol{k})\leq E_c(\boldsymbol{k})\leq E_d(\boldsymbol{k})\right]$. Moreover, the eigenvalues have to be determined iteratively, since $n_f$, $m_f$ and $n_d$ appearing in $\mathcal{H}(\boldsymbol{k},\sigma)$ have to satisfy a self-consistency condition. This condition can easily be formulated in the half-filled case, where---as it will be discussed below---the two lower bands with dispersion $E_a(\boldsymbol{k})$ and $E_b(\boldsymbol{k})$ are fully occupied and the two higher lying bands are empty.  
The conditions of self-consistency for $n_f$ and $m_f$ are
\begin{gather}
n_{f} =  \frac{1}{N}\sum_{\boldsymbol{k},\sigma}{}^{'}
\Big[\mathcal{T}_{31}^2(\boldsymbol{k}\sigma) + \mathcal{T}_{32}^2(\boldsymbol{k}\sigma)
+\mathcal{T}_{41}^2(\boldsymbol{k}\sigma) + \mathcal{T}_{42}^2(\boldsymbol{k}\sigma)\Big],\label{eq:self-cons1-AF}\\
m_f=\frac{4}{N}\sum_{\boldsymbol{k}}{}^{'}\big[\mathcal{T}_{31}(\boldsymbol{k}\uparrow)\mathcal{T}_{41}(\boldsymbol{k}\uparrow)+\mathcal{T}_{32}(\boldsymbol{k}\uparrow)\mathcal{T}_{42}(\boldsymbol{k}\uparrow)\big],\label{eq:self-cons2-AF}
\end{gather}
and the total energy is
\begin{gather}
\mathcal{E}_{\rm g}^{\rm AF}= 
\sum_{\boldsymbol{k},\sigma}{}^{'} \big[E_a(\boldsymbol{k})+E_b(\boldsymbol{k})\big]\ \ \ \ \ \ \ \ \ \ \ \ \ \ \ \ \ \ \ \ \ \ \ \ \ \ \ \ \ \nonumber\\
\ \ \ \ \ \ \ \ - \frac{NU_f}{4} \left(n_f^2-m_f^2\right) -NU_{df}n_fn_d.
\end{gather}

\begin{figure}[!htb]
\includegraphics[scale=0.5]{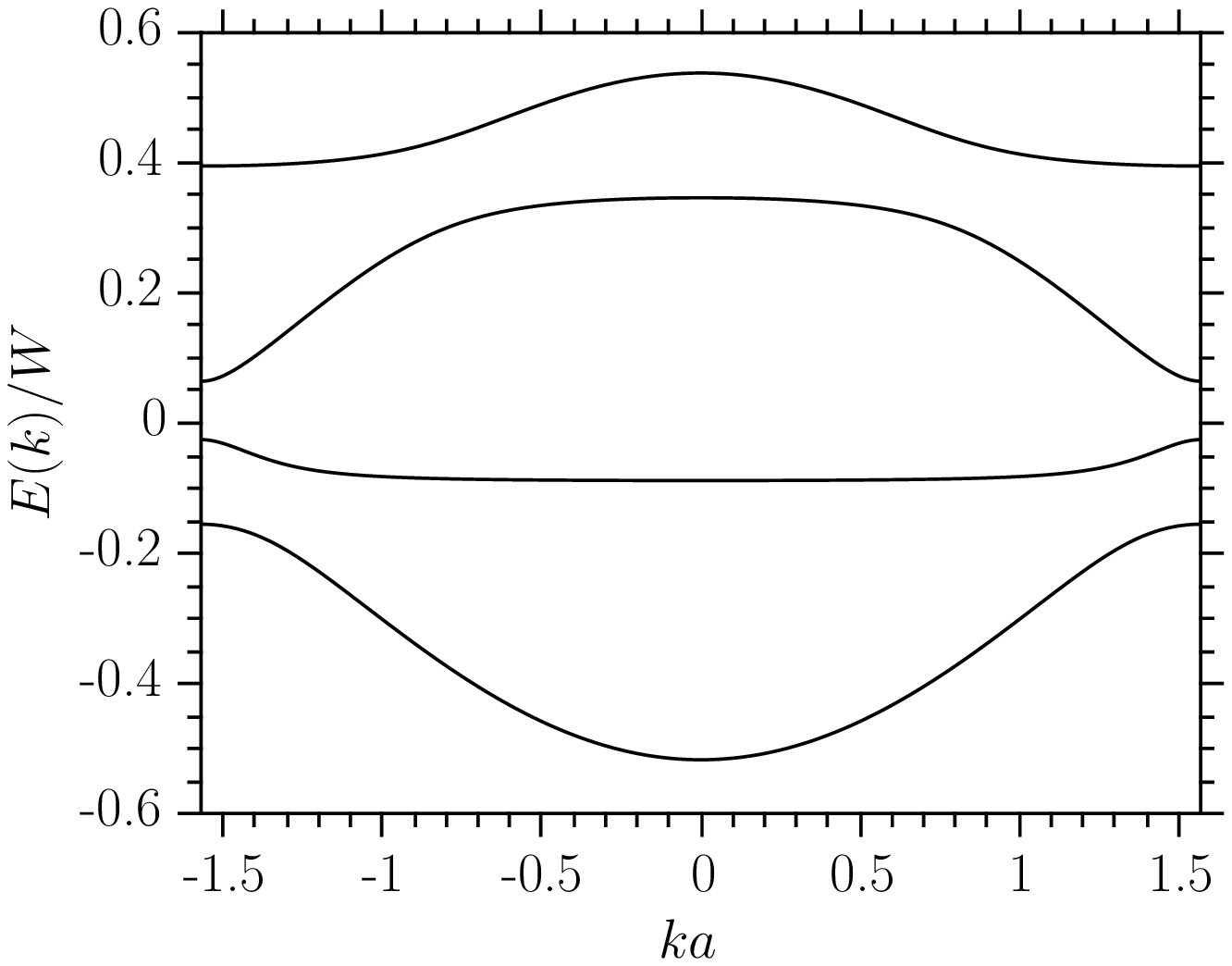}
\caption{\label{sav:fig}
The dispersion curves of the states diagonalizing the mean-field Hamiltonian for $U_f/W=0.5$, $U_{df}/W=0$, $\varepsilon_f/W=-0.1$, $V/W=0.1$.}
\end{figure}

Before evaluating the self-consistency equations, we return to the problem of the eigenvalue equations. Analytic expressions can be given at the symmetric point, and the obtained four bands for $U_{df}=0$ become simply 
\begin{gather}
E_{\alpha}(\boldsymbol{k}) = \pm \frac{1}{\sqrt{2}}\Bigg[\varepsilon_d^2(\boldsymbol{k}) + U_f^2m_f^2 +2V^2 \nonumber\\
\pm \left.\sqrt{\left(\varepsilon_d^2(\boldsymbol{k}) - U_f^2m_f^2 \right)^2 
+ 4V^2\left(\varepsilon_d^2(\boldsymbol{k}) + U_f^2m_f^2 \right)} \right]^{\frac{1}{2}}.
\end{gather}
The band structure is displayed in Fig. \ref{sav:fig} for the one-dimensional tight-binding case for a special, but non-symmetric choice of the parameters.
In higher dimensions, the system does not necessary remain insulating  if the gap opens at different energies in different points of the zone boundary.

The self-consistency equations always have a paramagnetic solution, $m_f=0$, and in certain cases they have a magnetic solution, $|m_f|>0$. We compare the energies of both solutions and accept that one, which has lower energy. It is worth noting that the polarization of the $d$-electrons, $m_d$, is also nonzero and its sign is opposite to $m_f$ in the magnetic solution (though its value is smaller than $m_f$ by an order of magnitude).

\begin{figure}[!htb]
\includegraphics[scale=0.5]{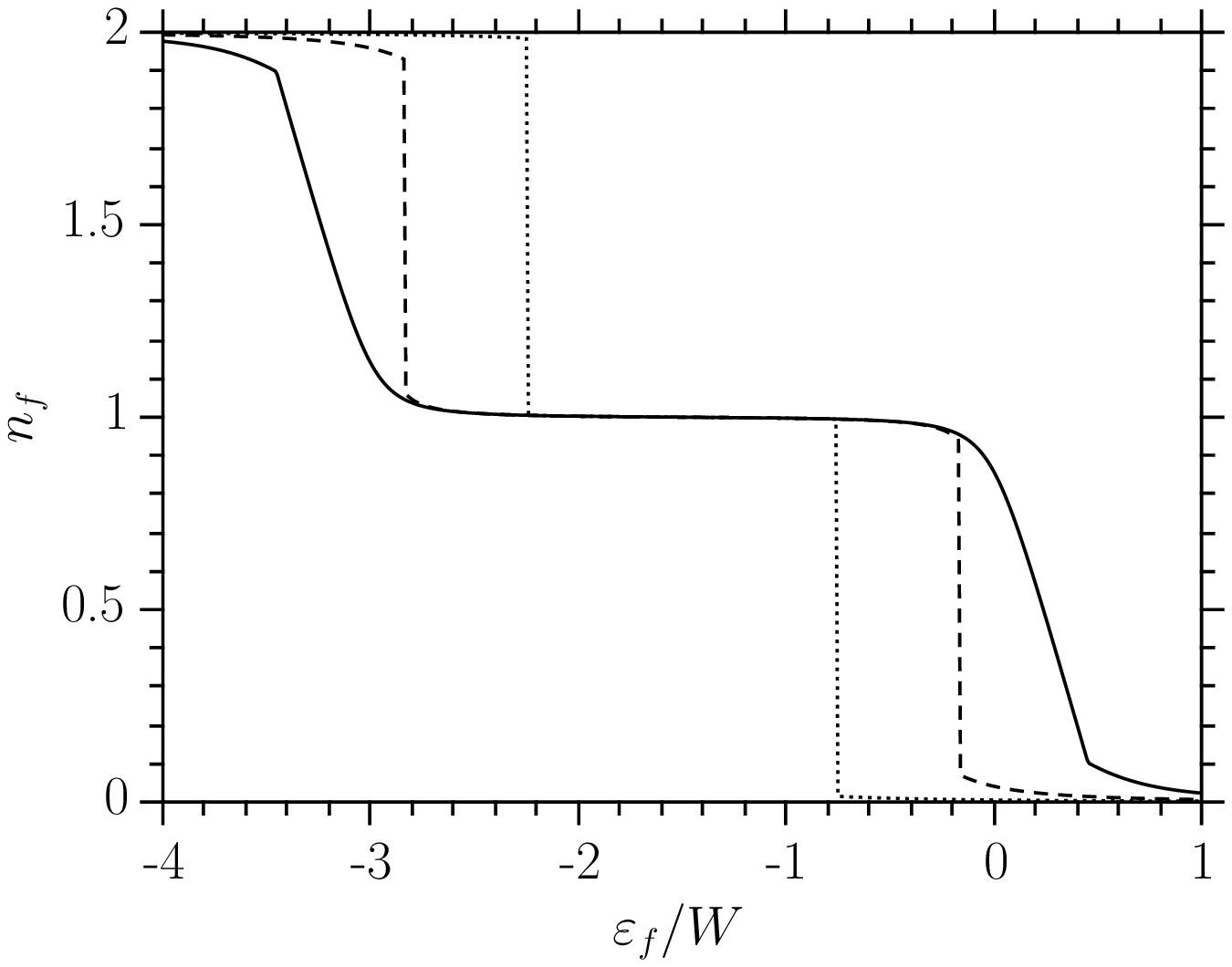}
\caption{\label{meanfield:fig}
The $f$-level occupancy as a function of $\varepsilon_f$ obtained by mean-field theory.  The solid, dashed and dotted lines belong to $U_{df}/W=0;0.4;1$, respectively, $V/W=0.1$, $U_f/W=3$ in all cases.}
\end{figure}

In the actual calculations we do not work in $\boldsymbol{k}$-space. The summations over $\boldsymbol{k}$ are carried out by assuming a constant density of states
$\rho(\varepsilon)=1/W$ in the interval $\varepsilon\in[-W/2,W/2]$. The same $\rho(\varepsilon)$ will be used in the variational calculation, too.
The numerical results for $n_f$ are shown in Fig. \ref{meanfield:fig} for different values of $U_{df}$. 
As long as the $f$-level is nearly fully occupied or empty, the paramagnetic solution is favorable, while when the occupancy is nearly 1, the magnetic solution has lower energy. For $U_{df}/W=0$, the $f$-level occupancy is continuous, although there is a 
discontinuity in its derivative at the point where the paramagnetic solution switches to magnetic or vice versa. This happens in Fig.\ \ref{meanfield:fig} at $\varepsilon_f/W\approx-3.45$ and 0.45.
For increasing $U_{df}$ values, first the mixed-valence regime narrows, then a
jump, a first-order transition appears in the $f$-level occupancy between the
paramagnetic and the magnetic solution and the Kondo regime shrinks rapidly.
This first-order transition has already been found by other
calculations.\cite{Hirashima:cikk,Kubo:GW} 
As is seen here, this simple mean-field approach can also account for it. 
The mean-field theory gives a critical value of $U_{df}^\text{c}/W\approx0.26$ for $V/W=0.1$, $U_f/W=3$. 

At a certain value of $U_{df}$ above $U_{df}^{\text{c}}$ the magnetic solution is stable in a single point, $\varepsilon_f=-U_f/2$. In this point the
paramagnetic solutions with $n_f\approx2$ and $n_f\approx0$, and the
magnetic solution with $n_f=1$ have the same energy, that is, three states
coexist.  This is a triple point, since three first-order transition lines
meet here. For larger $U_{df}$ values,
beyond the triple point, a so-called valence skipping occurs, that is, the
valence state $n_f\approx1$ is missing, since a direct first-order transition
takes place from $n_f\approx2$ to $n_f\approx0$. It is interesting to note that so far
the valence skipping (which was observed in several compounds, for
example BaBiO$_3$) has been attributed to the presence of a negative $U_f$.\cite{Varma:cikk} 
As it is demonstrated here, large enough $U_{df}$ can also
lead to valence skipping, even if $U_f>0$. The mean-field
theory gives $U_{df}^\text{triple}/W\approx1.75$ for
$V/W=0.1$, $U_f/W=3$. Note that this is in good agreement with the result $U_{df}^\text{triple}\approx U_f/2 +
W/4$ that will be obtained by the Gutzwiller method. 

This picture remains valid even when $U_f$ is small, but finite, compared to the bandwidth. Although no plateau with $n_f\approx1$ is formed, that is, there is no Kondo-regime, a stable magnetic solution is found near the symmetric point, and this regime is bounded by first-order transitions above a finite $U_{df}^{\text{c}}$. Note that for $U_f/W=0$ there is no triple
point due to the lack of magnetic solution. However, a direct first-order
transition appears between the paramagnetic states with
solution between  $n_f\approx2$ and $n_f\approx0$ above a certain value of $U_{df}$.

Although this theory gives a qualitatively good description of the first-order transition, there are several problems with it. Firstly, we had to assume a magnetic order of spin-density-wave type, while no long-range order is expected in the Kondo regime. Second, the valence susceptibility, which is defined by
\begin{gather} \label{eq:susceptibility}
 \chi_V=-\frac{dn_f}{d\varepsilon_f},
\end{gather}
 is not a continuous function, as is mentioned above, even below $U_{df}^c/W$, where $n_f$ is continuous. The mean-field approach does not provide us with a critical point, where $\chi_V$ would diverge. In order to find a QCP and to investigate its properties, we need a more accurate calculation. This is done in the next section by the variational method.

\section{Variational calculation}
In what follows we generalize the variational approach used in [Ref. \onlinecite{Itai:variational,Hagymasi:long}]  and summarize 
briefly the main steps of the calculation.
We restrict ourselves to the paramagnetic case, that is, the number of 
up-spin and down-spin electrons are assumed to be equal locally, too. As it will be pointed out, the quantum criticality and the first-order transition appear without any further assumptions, in contrast to the mean-field calculation. The trial state is expressed in terms of Gutzwiller projectors:
\begin{gather}
|\Psi\rangle = \hat{P}(f^1d^2)\hat{P}(f^1d^1)\hat{P}(f^2d^2)\hat{P}(f^2d^1)\hat{P}(f^2)|\Psi_0\rangle, 
\end{gather}
where
\begin{gather}
|\Psi_0\rangle=\prod_{\sigma}\prod_{\boldsymbol{k}}\left[u(\boldsymbol{k})\hat{f}_{\boldsymbol{k},\sigma}^{\dagger}+v(\boldsymbol{k})\hat{d}_{\boldsymbol{k},\sigma}^{\dagger}\right]|0\rangle
\end{gather}
contains the mixing amplitudes $u(\boldsymbol{k})$ and $v(\boldsymbol{k})$ as variational parameters,
and the sum over $\boldsymbol{k}$ extends over the whole Brillouin zone. The Gutzwiller projectors $\hat{P}(f^{\alpha}d^{\beta})$, where $\alpha$ and $\beta$ denote the $f$- and $d$-electron numbers respectively,  act on the on-site electron configurations defined by their arguments, making that configuration less probable. For example:
\begin{gather}
  \hat{P}(f^2)=\prod_{\boldsymbol{g}}\left[1-\left(1-\eta(f^2)\right)\hat{n}_{\boldsymbol{g}\uparrow}^f\hat{n}_{\boldsymbol{g}\downarrow}^f\right]
\label{eq:Gutzwiller_proj_f}
\end{gather}
is the standard Gutzwiller-projector for two $f$-electrons on the same site. The other ones take into account correlations between $d$- and $f$-electrons, for example:
\begin{gather}
 \hat{P}(f^1d^2)=\prod_{\boldsymbol{g}}\left\{1-\left(1-\eta(f^1d^2)\right)\left[\hat{n}_{\boldsymbol{g}\uparrow}^f\left(1-\hat{n}_{\boldsymbol{g}\downarrow}^f\right)\right.\right.\nonumber\\
+\left.\left.\hat{n}_{\boldsymbol{g}\downarrow}^f\left(1-\hat{n}_{\boldsymbol{g}\uparrow}^f\right)\right]\hat{n}_{\boldsymbol{g}\uparrow}^d\hat{n}_{\boldsymbol{g}\downarrow}^d\right\},\\
 \hat{P}(f^2d^2)=\prod_{\boldsymbol{g}}\left\{1-\left(1-\eta(f^2d^2)\right)\hat{n}_{\boldsymbol{g}\uparrow}^f\hat{n}_{\boldsymbol{g}\downarrow}^f\hat{n}_{\boldsymbol{g}\uparrow}^d\hat{n}_{\boldsymbol{g}\downarrow}^d\right\}.
\end{gather}
The remaining projectors are defined straightforwardly. Besides the mixing amplitudes, we have five variational parameters, $\eta(f^2d^2)$, $\eta(f^2d^1)$, $\eta(f^1d^2)$, $\eta(f^1d^1)$, 
$\eta(f^2)$, controlled by $U_{df}$ and $U_f$. The tedious procedure of
optimization is omitted here. 
Performing the optimization with respect to the mixing amplitudes using the Gutzwiller approximation we obtain
\begin{gather}
 \mathcal{E}=\frac{1}{N}\sum_{\boldsymbol{k}}\left[q_d\varepsilon_d(\boldsymbol{k})+
    \tilde{\varepsilon}_f-\sqrt{\big[q_d\varepsilon_d(\boldsymbol{k})
     -\tilde{\varepsilon}_f\big]^2+4\tilde{V}^2}\right]   \nonumber\\
      +(\varepsilon_f-\tilde{\varepsilon}_f)n_f+U_f\nu(f^2)\nonumber\\
      +U_{df}\left[4\nu(f^2d^2)+2\left(\nu(f^2d^1)+\nu(f^1d^2)\right)+\nu(f^1d^1)\right]
\label{eq:energy}
\end{gather}
for the ground-state energy density, where $\nu(f^2)$ is the density of 
doubly occupied $f$-sites. The other $\nu(f^{\alpha}d^{\beta})$ quantities denote the corresponding densities of the $f^{\alpha}d^{\beta}$ configurations, e.g.:
\begin{gather}
 \nu(f^1d^2)=\frac{1}{N}\left\langle\left[\hat{n}_{\boldsymbol{g}\uparrow}^f\left(1-\hat{n}_{\boldsymbol{g}\downarrow}^f\right)+
\hat{n}_{\boldsymbol{g}\downarrow}^f\left(1-\hat{n}_{\boldsymbol{g}\uparrow}^f\right)\right]\hat{n}_{\boldsymbol{g}\uparrow}^d\hat{n}_{\boldsymbol{g}\downarrow}^d \right\rangle,\nonumber\\
\nu(f^2d^2)=\frac{1}{N}\left\langle\hat{n}_{\boldsymbol{g}\uparrow}^f\hat{n}_{\boldsymbol{g}\downarrow}^f\hat{n}_{\boldsymbol{g}\uparrow}^d\hat{n}_{\boldsymbol{g}\downarrow}^d\right\rangle.
\end{gather}
$\tilde{V}=V\sqrt{q_fq_d}$ is the renormalized hybridization, $q_f$ and $q_d$ are the kinetic energy renormalization factors for the $f$- and $d$-electrons, respectively. Their analytic forms are now much longer than in our previous paper, \cite{Hagymasi:long} and after a tedious algebra we arrive at the following complete square forms: 
\begin{widetext}
 \begin{gather} 
  q_{f}=\frac{1}{(n_{f}/2)(1-(n_{f}/2))}\left(\sqrt{\nu(f^2d^2)\nu(f^{1}d^2)}+2\sqrt{\nu(f^2d^{1})\nu(f^{1}d^{1})}+\sqrt{\nu(f^{2}d^{0})\nu(f^{1}d^{0})}+\right.\nonumber\\
\label{eq:qf_full}\left.\sqrt{\nu(f^{1}d^{2})\nu(f^0d^{2})}+2\sqrt{\nu(f^{1}d^{1})\nu(f^0d^{1})}+\sqrt{\nu(f^{1}d^{0})\nu(f^{0}d^{0})}\right)^2,\\
q_{d}=\frac{1}{((n/2)-(n_{f}/2))(1-(n/2+n_{f}/2))}\left(\sqrt{\nu(f^2d^2)\nu(f^{2}d^{1})}+2\sqrt{\nu(f^{1}d^{2})\nu(f^{1}d^{1})}+\right.\nonumber\\
\left.\sqrt{\nu(f^{0}d^{2})\nu(f^{0}d^{1})}+\sqrt{\nu(f^2d^{1})\nu(f^{2}d^{0})}+2\sqrt{\nu(f^{1}d^{1})\nu(f^{1}d^{0})}+\sqrt{\nu(f^{0}d^{1})\nu(f^{0}d^{0})}\right)^2,
\label{eq:qd_full}
 \end{gather}
\end{widetext}
where $n$ is the band filling, which is 2 in our case.
It is remarkable that the renormalized hybridization can still be written as the square root of $q_f$ and $q_d$ in the presence of 
$U_{df}$, too. Furthermore, $\tilde{\varepsilon}_f$, the quasiparticle energy level of $f$-electrons, has the same form as 
in [Ref. \onlinecite{Hagymasi:long}]. It provides a self-consistency equation for $n_f$. 
The summation over $\boldsymbol{k}$ in Eq.\ (\ref{eq:energy}) is carried out
with a constant density of states, $\rho(\varepsilon)=1/W$, in the interval
$\varepsilon\in[-W/2,W/2]$. During the optimization process, the
$\eta(f^{\alpha}d^{\beta})$ variational parameters are expressed as functions
of the quantities $\nu(f^{\alpha}d^{\beta})$, therefore the actual optimization can
be done with respect to these parameters.
All in all, the energy density given in (\ref{eq:energy}) has to be optimized for $n_f$ and $\nu(f^2)$, $\nu(f^2d^1)$, $\nu(f^2d^2)$, $\nu(f^1d^1)$, $\nu(f^1d^2)$, the other quantities appearing in Eqs. (\ref{eq:qf_full})-(\ref{eq:qd_full}) can be expressed using these due to the conservation of the number of particles. The evaluation of the variational equations could be done only numerically.
\par
\begin{figure}[!htb]
\includegraphics[scale=0.5]{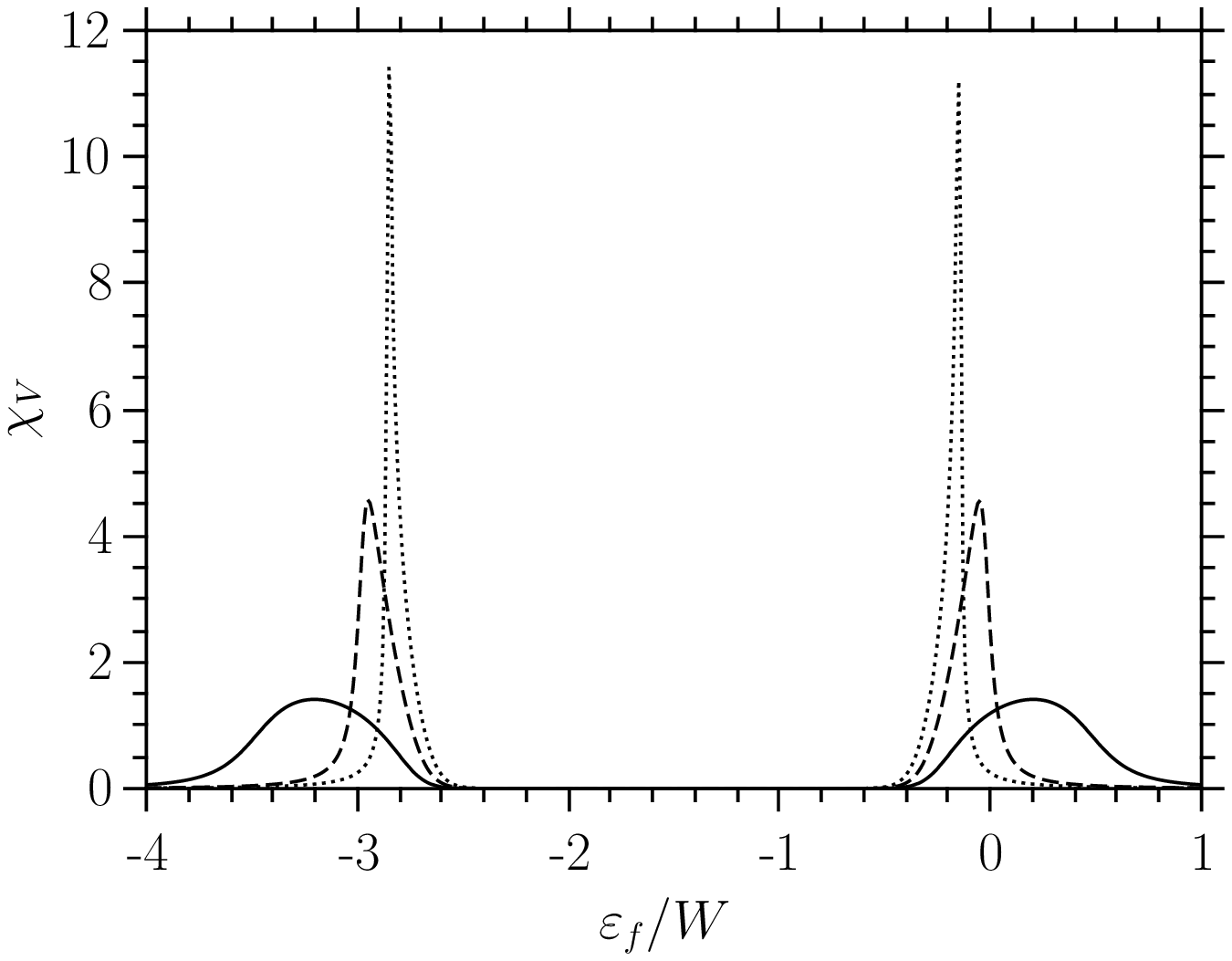}
\caption{\label{nf_gorbek:fig}
The valence susceptibility as a function of the $f$-level energy for $U_f/W=3$ and $V/W=0.1$. The solid, dashed and dotted lines correspond to $U_{df}/W=0;0.3$ and $0.4$, respectively.}
\end{figure}

 We first address what happens in the mixed-valence regime. As $U_{df}$ is switched on, the mixed-valence regimes tend to be sharper and sharper. This can be characterized by the valence susceptibility defined in Eq.\ (\ref{eq:susceptibility}) and displayed in Fig. \ref{nf_gorbek:fig} for different values of $U_{df}$. Note that in the half-filled case, this function is symmetric to the point $\varepsilon_f=-U_f/2$, therefore it is sufficient to investigate the critical behavior in the regime $0<n_f<1$. In what follows, we focus on this regime, if not mentioned otherwise.
It is found,  in agreement with other calculations,\cite{Hirashima:cikk,Kubo:GW} that $\chi_V$ diverges for a  certain value of $U_{df}^{\rm c}$ and two values of $\varepsilon_f^{\rm c}$ related by the symmetry with respect to $-U_f/2$. These two points are identified as the QCPs. Following the maximum values of $\chi_V$, the position of the QCP can be determined. We found that $\chi_V$ diverges as
\begin{gather}
 \chi_V|_{\varepsilon_f=\varepsilon_f^\text{c}}\sim \frac{1}{|U_{df}-U_{df}^\text{c}|},\\
 \chi_V|_{U_{df}=U_{df}^\text{c}}\sim \frac{1}{|\varepsilon_{f}-\varepsilon_{f}^\text{c}|}.
\end{gather}
This power-law behavior is valid for every choice of the parameters we used in our calculations, indicating universality. 
\par
\begin{figure}[!h]
\includegraphics[scale=0.5]{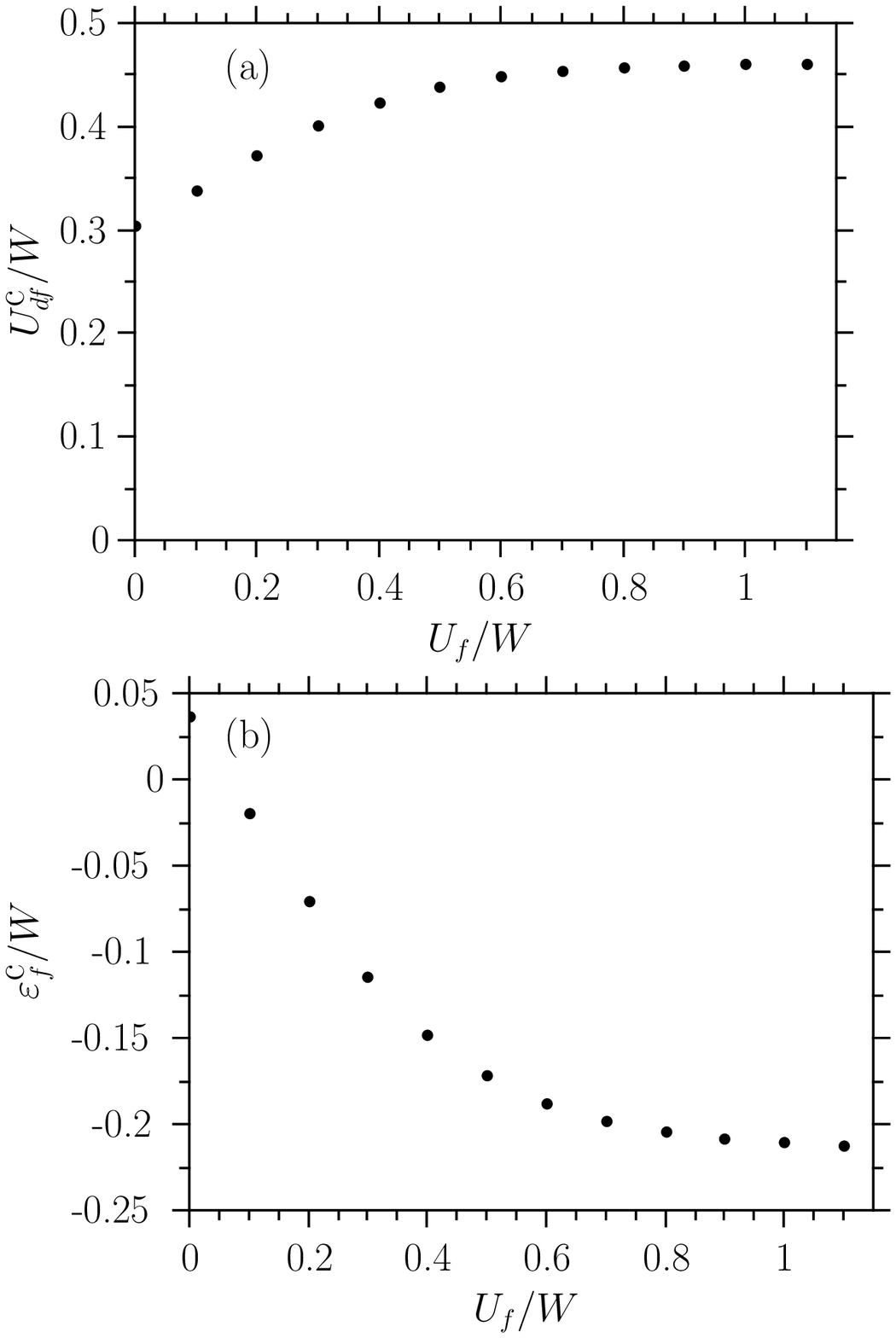}
\caption{\label{Uf_krit:fig}
Panel (a) shows the critical value of $U_{df}$ where the QCP appears as a function of $U_f$. Panel (b) shows the critical value of $\varepsilon_f$, $V/W=0.1$ in all cases. }
\end{figure}

\begin{figure}[!htb]
\includegraphics[scale=0.5]{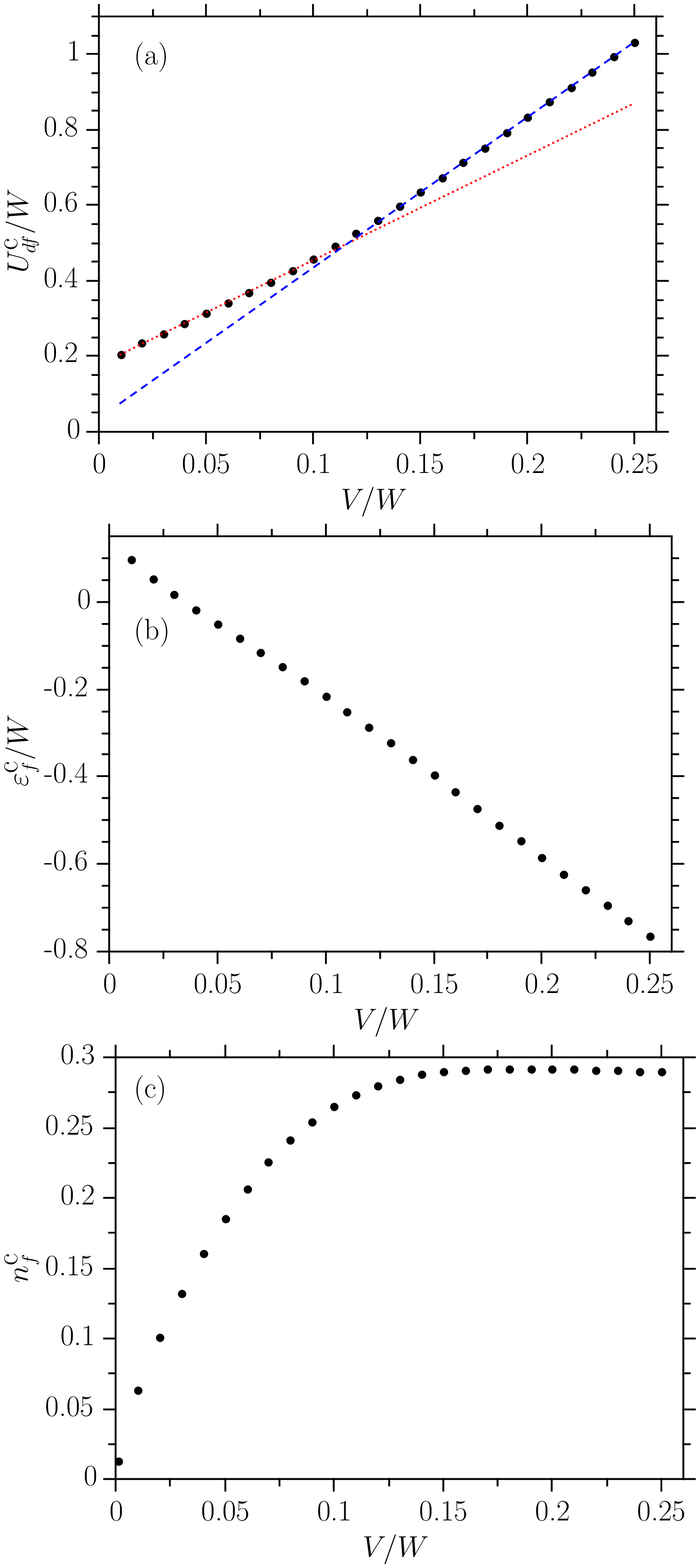}
\caption{\label{Udf_krit:fig}
(Color online) Panel (a) shows the critical value of $U_{df}$ (\textbullet) where the QCP appears as a function of the hybridization. The red dotted and blue dashed lines are linear fits to the beginning and to the end of the data respectively. Panel (b) and (c) show the critical values of $\varepsilon_f$ and $n_f$ respectively, $U_f/W=3$ in all cases. }
\end{figure}

 Our Gutzwiller calculation makes it possible to investigate how the position of the QCP depends on $U_f$ and $V$. 
In Fig. \ref{Uf_krit:fig} the critical $U_{df}^{\rm c}$ and
$\varepsilon_f^{\rm c}$ (for the QCP in the $0<n_f<1$ regime) are shown as a function of $U_f$ for a
fixed $V$. We found that (i) even for $U_f=0$ there exists a critical point; (ii)
$U_{df}^{\rm c}$ and $\varepsilon_f^{\rm c}$ vary monotonically as $U_f$
increases; (iii) both of them saturate as $U_f$ reaches the value above which
there exists a stable Kondo regime (see Fig.\ 3 in [Ref.\
\onlinecite{Hagymasi:long}]). 
On the contrary their dependence on $V$ is remarkable. These values are shown in Fig. \ref{Udf_krit:fig}. 
Firstly, we mention, that for $V/W\rightarrow0$, $U_{df}^\text{c}$ tends to a nonzero value ($U_{df}^\text{c}/W\approx0.17$), which indicates that there has to be a finite value of $U_{df}$ even for weak hybridizations to obtain a valence transition. 
The critical position of the $f$-level decreases linearly with increasing hybridization. The critical value of the occupancy of the 
$f$-level increases from $n_f^\text{c}=0$ at $V=0$ and saturates as soon as $\varepsilon_f^\text{c}$ reaches the bottom of the conduction band. Roughly at the same mixing $V$, the slope of the $U_{df}^\text{c}-V$ curve shows a substantial change.
\par For larger values of $U_{df}$, two subsequent first-order transitions---from
$n_f\approx2$ to $n_f\approx1$ (Kondo regime) and from $n_f\approx1$ to
$n_f\approx0$---take place as $\varepsilon_f$ is varied. Their positions are symmetric with respect to $-U_f/2$.
This is confirmed by the fact that near the
transition a hysteresis is observed, that is, there is a narrow range of
$\varepsilon_f$, where two solutions of the variational equations coexist.
Therefore the transition line is identified from the ground-state energy, where
the energies of the different configurations are equal. 
\par For even larger values
of $U_{df}$, the width of Kondo regime decreases and at $U_{df}^{\text{triple}}$
it ends in a triple point. At the triple point, the energy of the Kondo-like
state becomes equal to the energy of the states with $n_f\approx2$ and $n_f\approx0$,
therefore here three different states coexist. We found that the triple point is
located at $\varepsilon_f^{\text{triple}}=-U_f/2$ and
$U_{df}^{\text{triple}}\approx U_f/2+W/4$, if there is a Kondo plateau. 
For small $U_f$ (including $U_f=0$), when there is no Kondo plateau, the numerical results can be fitted to
$U_{df}^{\text{triple}}\approx U_f/2 + W/3$. In [Ref. \onlinecite{Hirashima:cikk}]
using DMFT it was found that the Kondo 
regime is stable for $U_{df}\lesssim U_f/2$. Our result is in agreement with
this. 
The mean-field theory gives similar results except for $U_f=0$, where the triple point does not exist.
\begin{figure}[!h]
\includegraphics[scale=0.4]{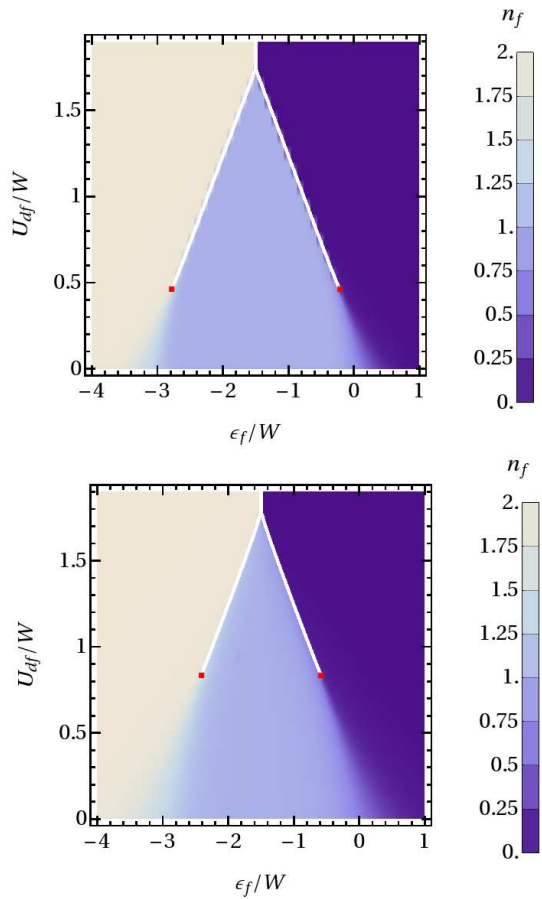}
\caption{\label{Udf_phase:fig}
(Color online) The upper panel shows the $\varepsilon_f-U_{df}$ phase diagram for $U_f/W=3$ and $V/W=0.1$, while the lower one for $U_f/W=3$ and $V/W=0.2$. The solid white lines denote the first-order transition lines. The QCPs are marked by red squares. The point where the white lines meet is the triple point (see the text).}
\end{figure}
Now we can draw the $\varepsilon_f-U_{df}$ phase diagram. The results are shown in Fig. \ref{Udf_phase:fig}, using a color code, for two different values of the hybridization and demonstrates our statements described above. The figure demonstrates that the interval of $\varepsilon_f$, where first-order transition occurs, is shortened for larger values of the hybridization.

\section{Comparison with the mean-field approach and exact diagonalization}
\begin{figure}[!htb]
\includegraphics[scale=0.5]{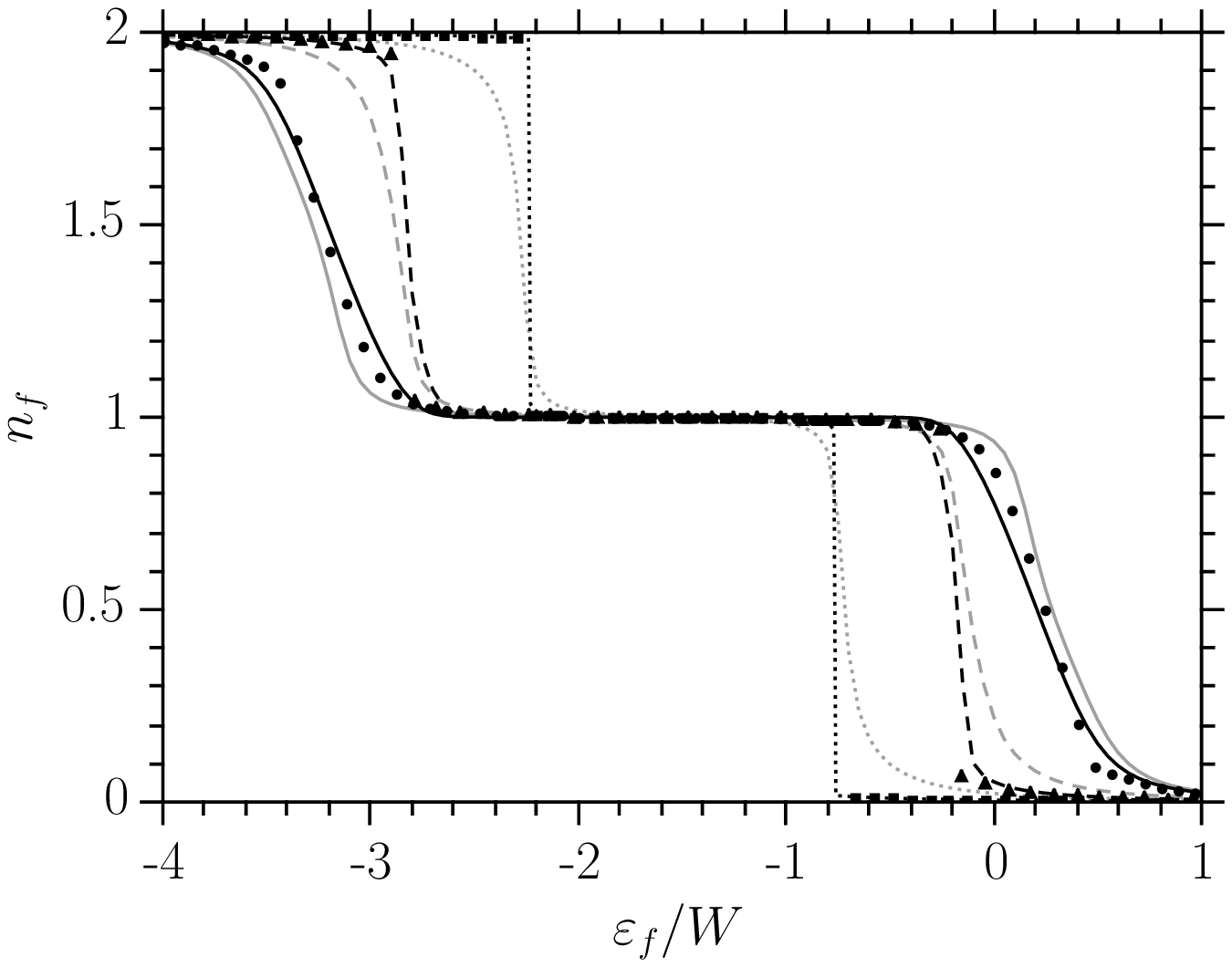}
\caption{\label{comparison:fig}
The $f$-level occupancy as a function of $\varepsilon_f$. The black curves are obtained from the Gutzwiller method, while the gray ones are calculated from exact diagonalization and the symbols are the results of the mean-field calculation. The solid, dashed and dotted lines (and the symbols \textbullet, $\blacktriangle$, $\blacksquare$) belong to $U_{df}/W=0;0.4;1$ respectively, $V/W=0.1$ and $U_f/W=3$ in all cases.}
\end{figure}

The mean-field theory and the Gutzwiller approach yield surprisingly close results for $U_{df}/W=0$ and $1$, which is shown in Fig. \ref{comparison:fig} for a special choice of the parameters. 
However, the mean-field results
show a jump in the $f$-level occupancy at such small values of $U_{df}$, where the Gutzwiller
method still gives a continuous change of $n_f$. 
The estimated critical value ($U_{df}^{\rm c}$) from
mean-field theory is significantly smaller than that from the Gutzwiller method.  
It is worth emphasizing that the jump in the mean-field
results is due to a level crossing between a paramagnetic and a magnetically ordered state. In
contrast, the Gutzwiller method gives a valence transition between paramagnetic
states. Both methods
result in a triple point for a certain value of $U_{df}$, and we
found that both of them gives $U_{df}^{\text{triple}}\approx U_f/2+W/4$. For
small values of $U_f$ the scenario is the same in both methods as in the large
$U_f$ case, however, the values of $U_{df}^{\text{triple}}$ and $U_{df}^{\rm c}$
are somewhat different. The only exception is
$U_f=0$, where there is no triple point in the mean-field calculation due to
the missing of a stable magnetic solution. Here a direct first-order transition takes place
between the nearly fully occupied and nearly empty $f$-levels.

As a further check of our results, we performed exact
diagonalization on a one-dimensional chain. Due to the limitation to relatively
short chains containing six sites, we do not expect to find critical behavior in
this calculation. However, some other features of the effect of $U_{df}$ might
be observable. The comparison is shown in Fig. \ref{comparison:fig}.
The width of the Kondo plateau is the same using all the three methods,
therefore its shrinking due to $U_{df}$ is not an artifact of the Gutzwiller approximation or
the mean-field treatment. Furthermore, by increasing $U_{df}$, the intermediate
valence regime becomes narrower in the exact diagonalization, too, although
there is naturally no sharp valence transition.

\section{Conclusions} We have performed mean-field calculation, variational calculation using the Gutzwiller method, and exact diagonalization for the extended PAM, where an additional local Coulomb
interaction between the $d$- and $f$-electrons has been included. Earlier calculations found a sharp, first-order valence transition and a critical point at some value of $U_{df}$ for large or
infinite $U_f$ couplings. We have generalized the Gutzwiller method for arbitrary $U_f$ in order to study the small $U_f$ regime and to analyze how the QCP depend on $U_f$ and $V$. 
\par Both the mean-field theory and the Gutzwiller method have resulted in 
two subsequent first-order valence transitions as the position of the $f$-level is varied
above a critical value  of $U_{df}$, and two QCPs appear in the $\varepsilon_f-U_{df}$ plane. We have analyzed variationally the critical behavior as a function of hybridization, the bare $f$-level
energy, and $U_f$, and have drawn the $\varepsilon_f-U_{df}$ phase diagram.  It has been pointed out that the Kondo regime shrinks by increasing $U_{df}$, and ends in a triple point, which obviously
cannot be seen in the infinite $U_f$ case. For even larger values of $U_{df}$ a direct first-order valence transition takes place from $n_f\approx2$ to $n_f\approx0$. This can be interpreted as
valence skipping, which so far has been attributed to the presence of a negative $U_f$. We find it for $U_f>0$, when $U_{df}$ is large enough. The shrinking of the Kondo regime and the narrowing of
the intermediate valence regime have been confirmed by exact diagonalization, although naturally, no sharp valence transition is found in finite chains.

\acknowledgments{This work was supported in part by the
Hungarian Research Fund (OTKA) through Grant No.~T 68340.}

\end{document}